\documentstyle[aps,prb,tighten,floats,epsfig]{revtex}

\begin{document}

\twocolumn[

\hsize\textwidth\columnwidth\hsize
\csname@twocolumnfalse\endcsname

\draft
\title{Comment on ``Symmetry properties of magnetization in\\the Hubbard model at finite temperature''}
\author{A.~Avella, F.~Mancini and D.~Villani$^\dag$}
\address{Universit\`a degli Studi di Salerno --- Unit\`a INFM di Salerno\\
Dipartimento di Scienze Fisiche ``E.~R.~Caianiello'', 84081 Baronissi,
Salerno, Italy}
\date{\today}
\maketitle

\begin{abstract}
The results of G.~Su and M.~Suzuki {[Phys.~Rev.~B \textbf{54}, 8291
(1996); \textit{ibidem} \textbf{57}, 13367 (1998)]} for the spin and
pseudo-spin symmetry properties of the Hubbard model are reexamined. We
point out that the exact relations they have found are valid down to
zero temperature and that their solutions for both spin and pseudo-spin
correlation functions are incorrect.
\end{abstract}

]

\narrowtext

In a recent paper G.~Su and M.~Suzuki\cite{one} analyze the spin
symmetry properties enjoyed by the Hubbard model in presence of an
homogeneous external magnetic field. They derive at finite temperatures
an exact relation connecting the spin correlation function to the
magnetization. Also, the authors claim to have found without any a
priori assumption at least one of the exact solutions for the
magnetization as function of the applied field. This result follows
previous works\cite{two,three} for the pseudo-spin counterpart where a
solution for the pseudo-spin correlation as a function of the filling
has been assessed.

In this Comment we clarify the issue of applicability of the exact
relations, by use of the equation of motion, showing that they are valid
also at zero temperature. Furthermore, we show that the pretended
solutions for both spin and pseudo-spin correlation functions are
incorrect.

The Hubbard model in presence of an homogeneous external magnetic field
$h$ reads as
\begin{eqnarray}
H&=&\sum_{ij}\left(t_{ij}-\mu\delta_{ij}\right)c^\dagger(i)c(j)
+U\sum_in_\uparrow(i)n_\downarrow(i)\nonumber\\
&-&h\sum_is_z(i) \label{eq1}
\end{eqnarray}
where $s_z(i)$ is the third component of the spin density operator.

Let us introduce the total spin operators
\begin{eqnarray}
S^+&=&\sum_ic^\dagger_\uparrow(i)c_\downarrow(i)\nonumber\\
S^-&=&\sum_ic^\dagger_\downarrow(i)c_\uparrow(i)\label{eq2}\\
S_z&=&\frac12\sum_i\left(c^\dagger_\uparrow(i)c_\uparrow(i)
-c^\dagger_\downarrow(i)c_\downarrow(i)\right)\nonumber
\end{eqnarray}
and the thermal retarded Green's function
\begin{eqnarray}
S^{+-}\left(t-t'\right)&=&
\left\langle{\cal R}\left[S^+(t)S^-\left(t'\right)\right]\right\rangle\nonumber\\
                       &=&
\frac{i}{2\pi}\int_{-\infty}^{+\infty}\!\!\!\!\!\! d\omega\,e^{-i\omega\left(t-t'\right)}S^{+-}(\omega)
\label{eq3}
\end{eqnarray}
By means of the Hamiltonian (\ref{eq1}) the spin operators satisfy the
Heisenberg equations
\begin{equation}
i\frac\partial{\partial t}S^\pm=\pm2hS^\pm \;\;\;\;\;\;\;\;\;\;
i\frac\partial{\partial t}S^z=0
\label{eq4}
\end{equation}
Then, we have
\begin{equation}
\frac1NS^{+-}(\omega)=\frac{2m}{\omega-2h+i\eta}
\label{eq5}
\end{equation}
where $N$ is the number of sites and $m$ is the magnetization per site
\begin{equation}
m=\frac1N\left\langle S_z\right\rangle
\label{eq6}
\end{equation}

In presence of an external magnetic field the spin symmetry is
explicitly broken, [cfr. Eq.~(\ref{eq4})], and the propagator
$S^{+-}(\omega)$ exhibits a massive collective mode\cite{four}
$\omega=2h$. When $h=0$ and $m\neq0$ the collective mode becomes
gapless, in accordance with the Goldstone theorem.

From Eq.~(\ref{eq5}), by standard methods, we obtain the spin
correlation function
\begin{equation}
\frac1N\left\langle S^+\left(t\right)S^-\left(t'\right)\right\rangle=
\frac{2m\,e^{-2ih\left(t-t'\right)}}{1-e^{-2\beta h}}
\label{eq7}
\end{equation}
where $\beta=1/\!k_{\mathrm{B}}T$. Similarly, we derive
\begin{equation}
\frac1N\left\langle S^-\left(t\right)S^+\left(t'\right)\right\rangle=
-\frac{2m\,e^{2ih\left(t-t'\right)}}{1-e^{2\beta h}}
\label{eq8}
\end{equation}
Let us note that (\ref{eq7}) and (\ref{eq8}) satisfy the \textit{KMS}
relation:
\begin{equation}
\left\langle S^+\left(t\right)S^-\left(t'+i\beta\right)\right\rangle=
\left\langle S^-\left(t'\right)S^+\left(t\right)\right\rangle
\label{eq9}
\end{equation}

In the static case Eqs.~(\ref{eq7}) and (\ref{eq8}) become
\begin{eqnarray}
\frac1N\left\langle S^+S^-\right\rangle&=& m\left[\coth\left(\beta
h\right)+1\right]\nonumber\\
\frac1N\left\langle S^-S^+\right\rangle&=& m\left[\coth\left(\beta
h\right)-1\right]
\label{eq10}
\end{eqnarray}
These are the exact relations derived by Su and Suzuki\cite{one};
however they are not restricted at finite temperature, but hold also at
$T=0$. In particular, for finite magnetic field
\begin{eqnarray}
\lim_{T\rightarrow0}\frac1N\left\langle S^+S^-\right\rangle&=&2m\nonumber\\
\lim_{T\rightarrow0}\frac1N\left\langle S^-S^+\right\rangle&=&0
\label{eq11}
\end{eqnarray}

On the basis of the relations (\ref{eq10}) Su and Suzuki\cite{one}
promote as one of the exact solutions for the magnetization as a
function of the applied magnetic field the following expression
\begin{equation}
m\left(h,T\right)=\frac n2\tanh\left(\beta h\right)
\label{eq12}
\end{equation}
where $n$ is the particle density. Also, they stress that other
solutions, if they exist, might have similar forms. At the opposite, the
solution depicted in (\ref{eq12}) is clearly wrong except for the
limiting case of half-filling and infinite $U$. Indeed, the pretended
solution (\ref{eq12}) can be falsified by looking at two exactly
solvable limits of the Hubbard model. That is, the noninteracting [i.e.
$U=0$] and atomic [i.e. $t_{ij}=0$] ones.

\begin{figure}[tb]
\begin{center}
\epsfig{file=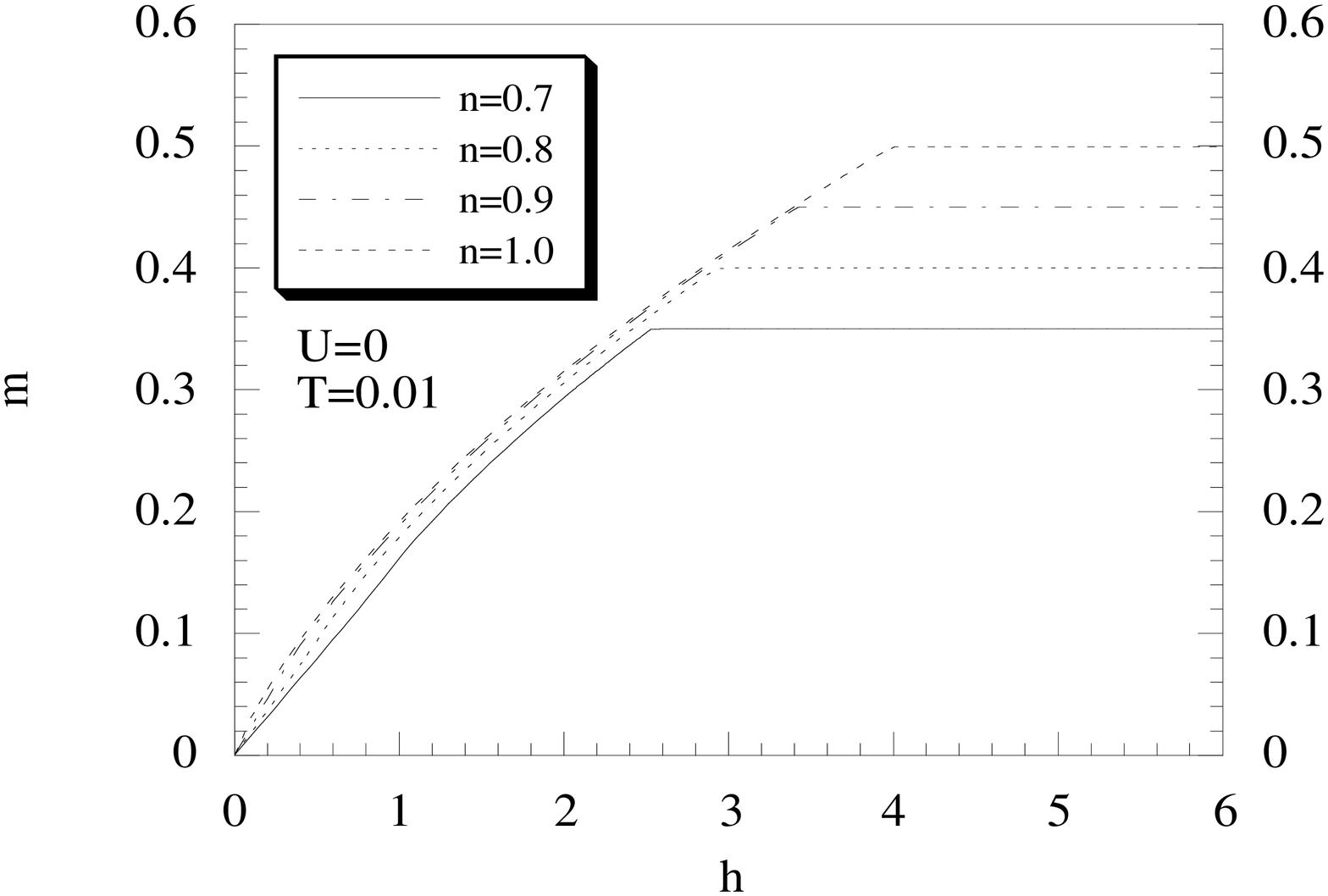,width=8cm,clip=}
\epsfig{file=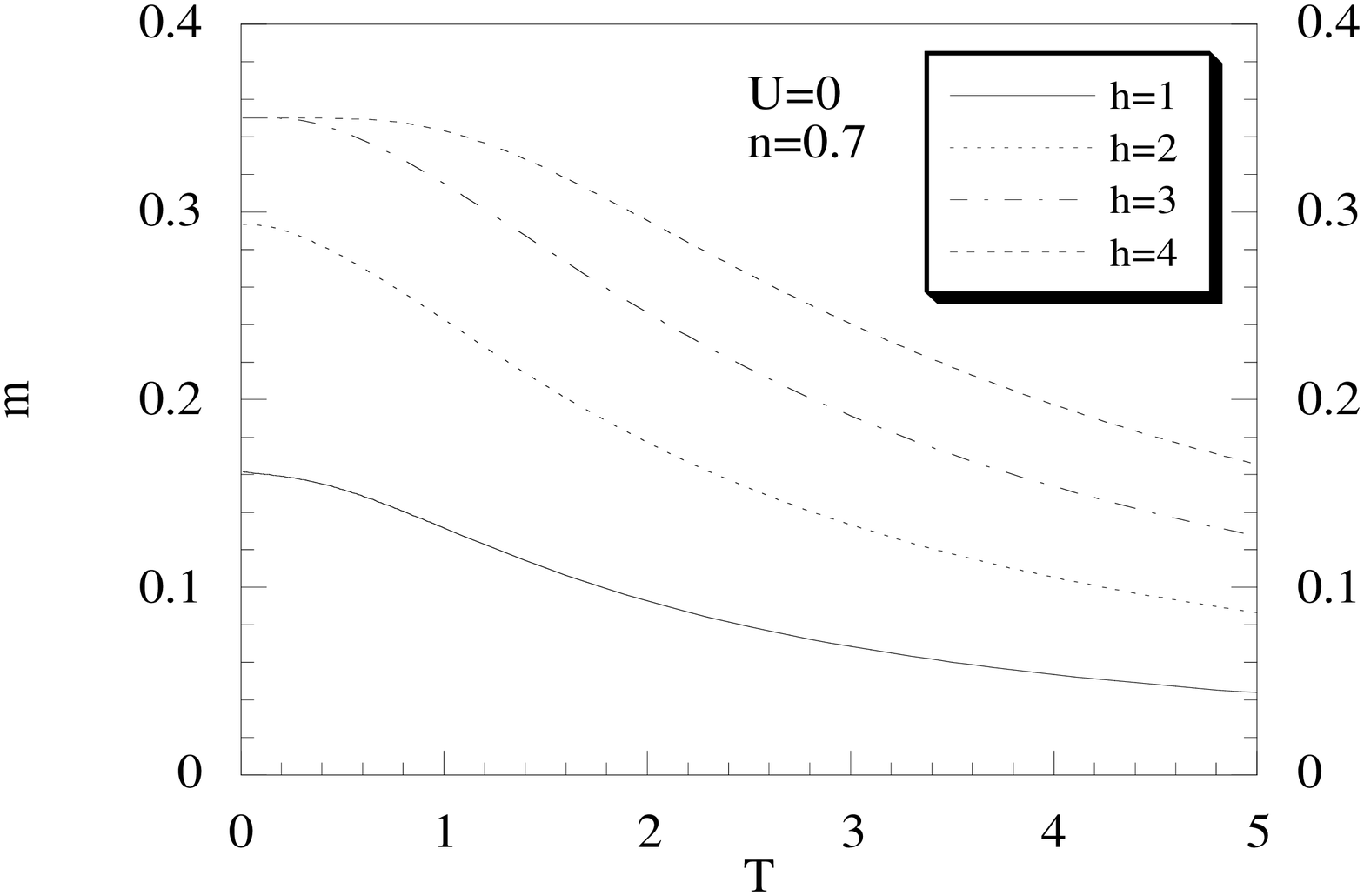,width=8cm,clip=}
\end{center}
\caption{Magnetization $m$ as a function of filling $n$,
applied magnetic field $h$ and temperature $T$.}
\label{fig1}
\end{figure}

\subsection*{Non interacting case}

It is direct to see that
\begin{eqnarray}
n&=& 1-\frac\Omega{2\left(2\pi\right)^d}
\int_{\Omega_{\mathrm{B}}} \!\!\! d^dk\left[T_\uparrow({\bf k})+T_\downarrow({\bf k})\right]
\label{eq13}\\
m&=&\frac14\tanh\left(\beta
h\right)\left[1-\frac\Omega{\left(2\pi\right)^d}
\int_{\Omega_{\mathrm{B}}} \!\!\! d^dk\,T_\uparrow({\bf k})T_\downarrow({\bf k})\right]
\label{eq14}
\end{eqnarray}
$\Omega$ is the volume of the unit cell, $d$ is the dimensionality of
the system and $\Omega_{\mathrm{B}}$ is the first Brillouin zone. We put
$T_\sigma({\bf k})=\tanh\left[\beta\,E_\sigma\right({\bf k})/2]$ with
energy spectra
\begin{eqnarray}
E_\uparrow({\bf k})&=&-\mu-4t\alpha({\bf k})-h \nonumber\\
E_\downarrow({\bf k})&=&-\mu-4t\alpha({\bf k})+h
\label{eq15}
\end{eqnarray}
where \mbox{$\alpha({\bf k})=1/\!d\sum_{i=1}^d\cos\left(k_ia\right)$},
$t$ is the hopping integral and $a$ is the lattice constant.

In the two-dimensional case the magnetization is shown (cfr.
Fig.~\ref{fig1}) for different values of the parameters $n$, $T$ and
$h$. The solution (\ref{eq12}), proposed in Ref.~\onlinecite{one},
obviously is not a solution for the case $U=0$.

\subsection*{Atomic limit}

In this case it is easy to show that
\begin{eqnarray}
n&=&\frac{1-\frac12\left(T_1+T_3\right)+\frac14\left(1-T_3\right)\left(T_1-T_2\right)}
{1-\frac14\left(T_1-T_2\right)\left(T_3-T_4\right)}\nonumber\\
&+&\frac{\frac14\left(1-T_1\right)\left(T_3-T_4\right)}
{1-\frac14\left(T_1-T_2\right)\left(T_3-T_4\right)}
 \label{eq16}\\
2m&=&\frac{\frac12\left(T_3-T_1\right)+\frac14\left(1-T_3\right)\left(T_1-T_2\right)}
{1-\frac14\left(T_1-T_2\right)\left(T_3-T_4\right)}\nonumber\\
&-&\frac{\frac14\left(1-T_1\right)\left(T_3-T_4\right)}
{1-\frac14\left(T_1-T_2\right)\left(T_3-T_4\right)}
\label{eq17}
\end{eqnarray}
where $T_i=\tanh\left(\beta\,E_i/2\right)$ with energy spectra
\begin{eqnarray}
E_1&=&-\mu-h \nonumber\\ E_2&=&-\mu+U-h\nonumber\\ E_3&=&-\mu+h
\nonumber\\ E_4&=&-\mu+U+h
\label{eq18}
\end{eqnarray}
We note that for $\mu=U/2$ the previous expressions become
\begin{eqnarray}
n&=&1\nonumber\\
m&=&-\frac{T_1+T_2}{4\left[1+\frac12\left(T_1-T_2\right)\right]}
=\frac{T_h\left(1+T_U\right)}{2\left(1+T_UT^2_h\right)}
\label{eq19}
\end{eqnarray}
with
\begin{equation}
T_U=\tanh\left(\beta\,U/\!4\right) \;\;\;\;\;\;\;\;
T_h=\tanh\left(\beta\,h/2\right)
\label{eq20}
\end{equation}

In particular, in the limit of large $U$ and for finite temperature
\begin{equation}
m\rightarrow\frac{T_h}{1+T^2_h}=\frac12\tanh\left(\beta\,h\right)
\label{eq21}
\end{equation}
in agreement with the result of Ref.~\onlinecite{five}.

An intrinsic symmetry of the Hubbard model is the pseudo-spin $SU(2)$
symmetry\cite{six} that combined with the spin $SU(2)$ one yields the
$[SU(2)\otimes SU(2)]/\!Z_2=SO(4)$ symmetry group. The generators of
this transformation are given by the total pseudo-spin operators
\begin{eqnarray}
P^+&=&\sum_ie^{i{\bf Q}\cdot{\bf
R}_i}c^\dagger_\uparrow(i)c^\dagger_\downarrow(i)\nonumber\\
P^-&=&\sum_ie^{-i{\bf Q}\cdot{\bf
R}_i}c_\downarrow(i)c_\uparrow(i)\label{eq22}\\
P_z&=&\frac12\sum_i\left[n(i)-1\right]\nonumber
\end{eqnarray}
where ${\bf Q}=(\pi,\pi)$. These operators satisfy the Heisenberg
equations
\begin{equation}
i\frac\partial{\partial t}P^\pm=\pm(2\mu-U)P^\pm \;\;\;\;\;\;\;\;\;\;
i\frac\partial{\partial t}P_z=0
\label{eq23}
\end{equation}

Let us consider the thermal retarded Green's function
\begin{eqnarray}
P^{+-}\left(t-t'\right)&=&
\left\langle{\cal R}\left[P^+(t)P^-\left(t'\right)\right]\right\rangle\nonumber\\
                       &=&
\frac{i}{2\pi}\int_{-\infty}^{+\infty}\!\!\!\!\!\! d\omega\,e^{-i\omega\left(t-t'\right)}P^{+-}(\omega)
\label{eq24}
\end{eqnarray}
By means of the equation of motion (\ref{eq23}) we find
\begin{equation}
\frac1NP^{+-}(\omega)=\frac{n-1}{\omega-(2\mu-U)+i\eta}
\label{eq25}
\end{equation}
The Hamiltonian has pseudo-spin $SU(2)$ symmetry only at half-filling;
when $n\neq1$ the symmetry is explicitly broken. Again, this is seen in
the propagator where a massive collective mode $\omega=2\mu-U$ is
observed\cite{four}. From (\ref{eq25}) we obtain the correlation
function
\begin{equation}
\frac1N\left\langle P^+\left(t\right)P^-\left(t'\right)\right\rangle=
\frac{(n-1)\,e^{-i(2\mu-U)\left(t-t'\right)}}{1-e^{-\beta(2\mu-U)}}
\label{eq26}
\end{equation}
and similarly
\begin{equation}
\frac1N\left\langle P^-\left(t\right)P^+\left(t'\right)\right\rangle=
-\frac{(n-1)\,e^{i(2\mu-U)\left(t-t'\right)}}{1-e^{\beta(2\mu-U)}}
\label{eq27}
\end{equation}
It is easy to see that (\ref{eq26}) and (\ref{eq27}) satisfy the
\emph{KMS} relation. In the static case:
\begin{eqnarray}
\frac1N\left\langle P^+P^-\right\rangle&=&
\frac12(n-1)\left[\coth\left(\beta(2\mu-U)/2\right)+1\right]\nonumber\\
\frac1N\left\langle P^-P^+\right\rangle&=&
\frac12(n-1)\left[\coth\left(\beta(2\mu-U)/2\right)-1\right]
\label{eq28}
\end{eqnarray}
which give
\begin{equation}
\left\langle P^+P^-\right\rangle=e^{\beta(2\mu-U)}\left\langle P^-P^+\right\rangle
\label{eq29}
\end{equation}

\begin{figure}[tb]
\begin{center}
\epsfig{file=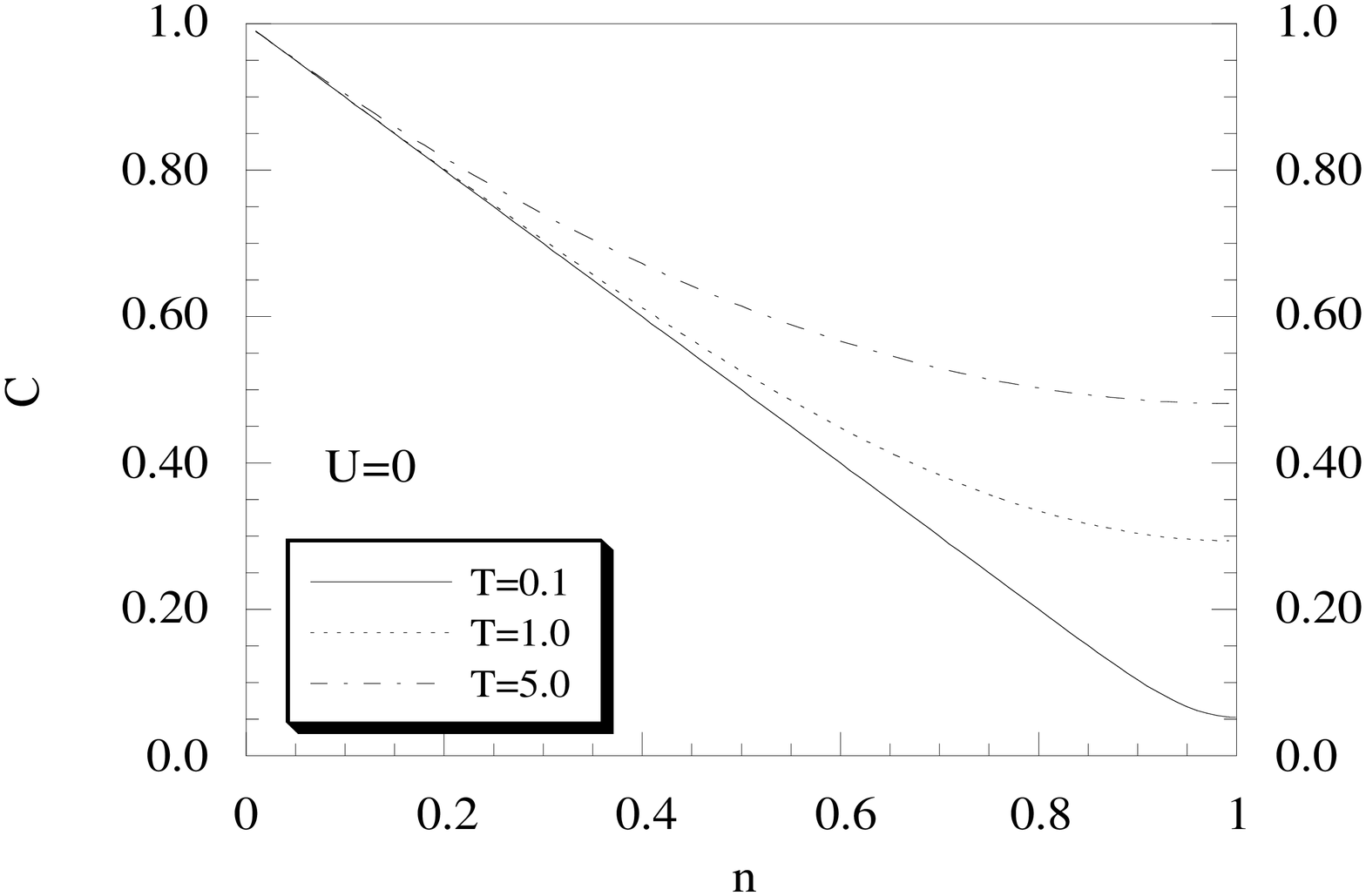,width=8cm,clip=}
\epsfig{file=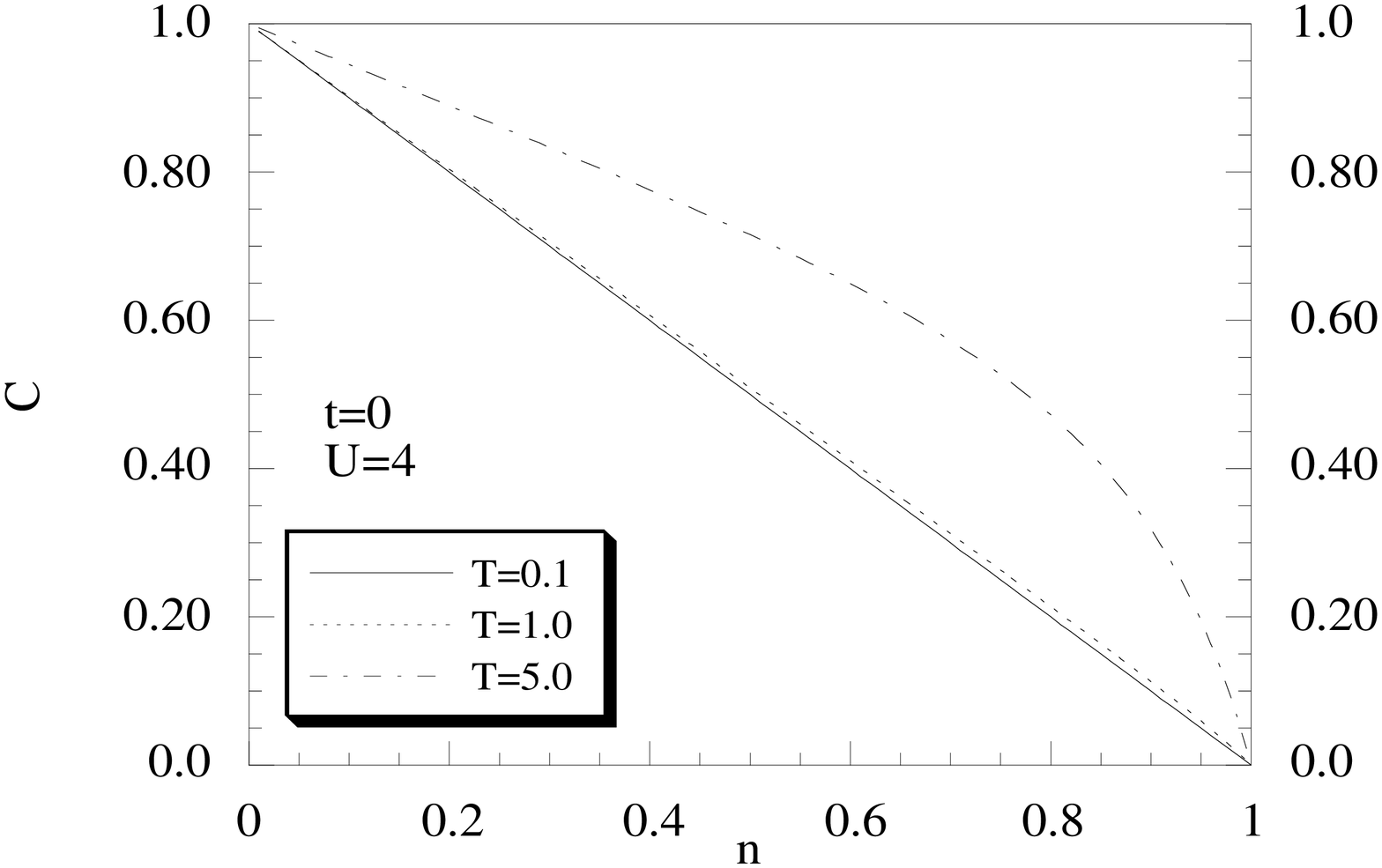,width=8cm,clip=}
\end{center}
\caption{$C$ as a function of filling $n$,
intrasite Coulomb repulsion $U$ and temperature $T$.}
\label{fig2}
\end{figure}

These exact results relates the pseudo-spin correlation functions to the
particle number $n$ and are a manifestation of the intrinsic symmetry.
These relations generalize at $T=0$ the results previously obtained by
Su\cite{three}.

Under the particle-hole transformation we have the following relations
\begin{equation}
\mu(2-n)=U-\mu(n) \;\;\;\;\;\;\;\; C^{+-}(2-n)=C^{-+}(n)
\label{eq30}
\end{equation}
where we put $C^{+-}=\left\langle P^+P^-\right\rangle$,
$C^{-+}=\left\langle P^-P^+\right\rangle$. Use of Eq.~(\ref{eq29}) leads
to
\begin{equation}
C^{+-}(2-n)=e^{-\beta\left[2\mu(n)-U\right]}C^{+-}(n)
\label{eq31}
\end{equation}
By making use of the transformation properties (\ref{eq30}), it is easy
to see that the expression (\ref{eq28}) satisfies the property
(\ref{eq31}). Actually, this is a manifestation of the intimate
interrelation between pseudo-spin and particle-hole
symmetries\cite{seven}.

In Ref.~\onlinecite{three} the particle-hole symmetry is tautologically
used as a supplementary equation and the following solution for the
pseudo-spin correlation function is presented for the case $h=0$
\begin{equation}
\frac1N\left\langle P^+P^-\right\rangle=\frac{C(T)}{1+e^{-\beta(2\mu-U)}}
\label{eq32}
\end{equation}
where $C(T)$ is an unknown function of temperature only. When
(\ref{eq32}) is used in (\ref{eq28}) one is lead to the following
equation for the chemical potential
\begin{equation}
n=1+C(T)\tanh\left[\beta(\mu-U2)\right]
\label{eq33}
\end{equation}
This equation is incorrect. For example, let us consider the limit of
small temperature. Then, (\ref{eq33}) would give $n\rightarrow1-C(0)$,
which is clearly wrong.

In Fig.~\ref{fig2} we present the function
$C(n,T,U)=(n-1)\coth\left[\beta(\mu-U2)\right]$ as a function of $n$ for
various temperatures, in the non-interacting and atomic limits. It is
clear that $C$ is not a function of temperature only, as stated in
Ref.~\onlinecite{three}, but varies with $n$ and $U$.

In conclusion, we have shown that the symmetry properties (\ref{eq10})
and (\ref{eq28}), obtained in Refs.~\onlinecite{one,two,three}, can be
derived by means of the equation of motion and are valid for any
temperature, including also zero temperature. These relations are exact
relations and are valid for any dimension of the system; for any value
of the Hubbard interaction $U$ and of the applied magnetic field $h$.
Furthermore, they relate bosonic and fermionic propagators. Indeed, any
approximation method should satisfy them in order to treat on equal
footing one- and two- particle Green's functions, and to preserve spin
and pseudo-spin symmetries\cite{eight}. We have also shown that the
solutions proposed in Refs.~\onlinecite{one} and \onlinecite{three} for
the magnetization and for the particle density are not valid.

\end{document}